\documentclass[12pt]{iopart}
\usepackage{iopams,bm,verbatim}
\begin{document}

\title{Newman Tamburino solutions with an aligned Maxwell field.}
\author{Liselotte De Groote and Norbert Van den Bergh }

\address{Department of Mathematical Analysis IW16, Ghent University, Galglaan 2, 9000 Ghent, Belgium}

\begin{abstract}
We prove that there exists no aligned Einstein Maxwell generalization of the spherical class of Newman Tamburino solutions. The presence of an aligned Maxwell field automatically leads to the cylindrical class.
\end{abstract}

\section{Introduction}
In 1962 Newman and Tamburino published the general empty space solutions for the class of metrics containing hypersurface orthogonal geodesic rays with nonvanishing shear and divergence~\cite{Newman1}. Their solutions can be subdivided into two classes, called `cylindrical' and `spherical', according to whether respectively $\rho^2-\sigma\,\bar{\sigma}=0$ or $\neq 0$.
In this paper we look at spherical solutions in the presence of an aligned Maxwell field. We will show that no such solutions exist.
\par
To prove this we use the Newman Penrose (NP) formalism. The choice of tetrad and the main equations are discussed in \S 2, while in \S 3 we prove our statement. The notations and conventions of \cite{Kramer} are followed throughout.

\section{Main equations}
Using a complex null tetrad $\left( \mathbf{m}, \mathbf{\bar{m}}, \mathbf{l}, \mathbf{k} \right)$ a solution of the Einstein-Maxwell equations is described in  the NP formalism by means of a) the 12 spin coefficients, which are independent complex linear combinations of the connection coefficients, b) the tetrad components of the Weyl tensor, which are expressed in terms of the five complex scalars
\begin{eqnarray}
\nonumber
\Psi_0 = C_{abcd}k^a m^b k^c m^d,  \quad \Psi_1 = C_{abcd} k^a l^b k^c m^d,   \\
\nonumber
\Psi_2 = \frac{1}{2} C_{abcd} k^a l^b \left( k^c l^d - m^c \bar{m}^d \right), \Psi_3 = C_{abcd} l^a k^b l^c \bar{m}^d, \Psi_4 = C_{abcd} l^a \bar{m}^b l^c \bar{m}^d
\end{eqnarray}
and c) the components
\begin{equation*}
\Phi_0= F_{ab} k^a m^b,\quad \Phi_1= \frac{1}{2} F_{ab} \left( k^a l^b + \bar{m}^a m^b \right),\quad \Phi_2= F_{ab} \bar{m}^a l^b
\end{equation*}
of the electromagnetic field. The Ricci tensor components are given by $\Phi_{ \alpha \beta } = \kappa_0 \Phi_{\alpha } \bar{\Phi}_{\beta}$ $\alpha , \beta = 0, 1, 2$.
The spin coefficients and curvature components satisfy the NP and Bianchi equations, which will be referred to as $NP_1, \ldots, NP_{18}$ and $B_1,\ldots, B_{11}$, following the same ordering as in~\cite{Kramer}.

\par
The variables $\Phi_0,\Phi_1,\Phi_2$ satisfy the Maxwell equations
\begin{eqnarray}
D \Phi_1 - \bar{\delta} \Phi_0 = \left( \pi - 2\, \alpha \right) \Phi_0 + 2\, \rho \Phi_1 - \kappa \Phi_2, \label{max1} \\
D \Phi_2 - \bar{\delta} \Phi_1 = - \lambda \Phi_0 + 2\, \pi \Phi_1 + \left( \rho - 2\, \epsilon \right) \Phi_2, \label{max2} \\
\delta \Phi_1 - \Delta \Phi_0 = \left( \mu - 2\, \gamma \right) \Phi_0 + 2\, \tau \Phi_1 - \sigma \Phi_2, \label{max3} \\
\delta \Phi_2 - \Delta \Phi_1 = - \nu \Phi_0 + 2\, \mu \Phi_1 + \left( \tau - 2\, \beta \right) \Phi_2. \label{max4},
\end{eqnarray}
under which three of the Bianchi equations ($B_9, B_{10},B_{11}$) become identities. 

As we will often make use of the commutators $\left[ \mathbf{e}_a , \mathbf{e}_b \right] = - 2\, \Gamma^c _{\left[ ab \right]} \mathbf{e}_c$, we present these explicitly in the appendix.

\section{Proof of the main result}

The Newman Tamburino metrics are characterized by the existence of a hypersurface orthogonal and geodesic principal null direction $\textbf{k}$, for which the shear and divergence are non-vanishing. In terms of the NP variables this translates into
\begin{equation}
\kappa = \epsilon + \bar{\epsilon} = 0, \label{kappanul}
\end{equation}
\begin{equation}
\Psi_0 = 0, \label{psianul}
\end{equation}
\begin{equation}
\rho - \bar{\rho} = 0. \label{rho}
\end{equation}
The spherical class of Newman Tamburino metrics, which we consider in the present paper, is further characterized by
\begin{equation}
\rho^2 - \sigma \, \bar{\sigma} \neq 0. \label{cyl}
\end{equation}
If this last equation (\ref{cyl}) does not hold, we are in the so called cylindrical class.
We suppose that the Maxwell field is \emph{aligned}, in the sense that $\mathbf{k}$ is not only a principal null vector of the Weyl tensor (which is expressed by (\ref{psianul})), but also that $\mathbf{k}$ is a principal null vector of the Maxwell tensor, which means that we can put $\Phi_0$ equal to 0 too.
\par

By means of a rotation, a boost and a null rotation of the null tetrad $\left(\mathbf{m}, \bar{\mathbf{m}}, \mathbf{l}, \mathbf{k} \right)$, we can make sure that the tetrad is parallelly propagated along the geodesic null congruence $\mathbf{k}$ :
\begin{equation}
\label{kappa}
\kappa = \epsilon = \pi = 0.
\end{equation}
Hereby the remaining boost freedom $\mathbf{k}\rightarrow A \mathbf{k}$ is restricted to boosts having $DA=0$. The NP equations guarantee then the existence of a solution of the equations $\delta  A = (\tau - \beta - \bar{\alpha})A$, $DA=0$, implying that
\begin{equation}
\label{tau}
\tau = \bar{\alpha} + \beta.
\end{equation}
\par
From the NP equations it also follows that $D(\sigma / \bar{\sigma})=0$ and so a final rotation $\mathbf{m}\rightarrow e^{i\theta} \mathbf{m}$ exists  such that
\begin{equation}
\label{sigma}
\bar{\sigma} = \sigma.
\end{equation}
The remaining tetrad freedom consists now of boosts $A$ and null rotations $B$ about $\mathbf{k}$, satisfying
\begin{equation*}
D\,A = \delta \, A  = \bar{\delta} \, A  = 0 = D \, B.
\end{equation*}

\par
Taking into account the above transformations (\ref{kappa}), (\ref{tau}) and (\ref{sigma}), we can rewrite the NP equations determining the evolution along the geodesic rays as follows:
\begin{equation*}
D\, \sigma  =2\,\rho\,\sigma ,
\end{equation*}
\begin{equation*}
D\, \rho ={\rho}^{2}+{\sigma}^{2} ,
\end{equation*}
\begin{equation*}
D\,\alpha =\rho\,\alpha+\beta\,\sigma ,
\end{equation*}
\begin{equation*}
D\, \lambda =\rho\,\lambda+\mu\,\sigma ,
\end{equation*}
\begin{equation*}
D\, \mu =\mu\,\rho+\sigma\,\lambda+\Psi_{{2}}+\frac{1}{12}\,R ,
\end{equation*}
\begin{equation*}
D\, \nu =\left( \alpha+\bar{\beta}\right)\mu +\left( \bar{\alpha} +\beta \right) \lambda +\Psi_{{3}}+ \Phi_2 \,\bar{\Phi_1} ,
\end{equation*}
\begin{equation*}
D\, \beta =\rho\,\beta+\sigma\,\alpha+\Psi_{{1}} ,
\end{equation*}
\begin{equation*}
D\, \gamma =\alpha\,\bar{\alpha} +2\,\alpha\,\beta+\beta\,\bar{\beta} +\Psi_{{2}}-\frac{1}{24}\,R+\Phi_1\, \bar{\Phi_1}.
\end{equation*}
We can also rewrite the Maxwell equations (\ref{max1}) and (\ref{max3}) :
\begin{equation*}
{D} \, \Phi_1 =2\,\rho\,\Phi_1 ,
\end{equation*}
\begin{equation}
\delta \, \Phi_1 =2 \left( \bar{\alpha} +\beta \right) \Phi_1 -\sigma\,\Phi_2 ,
\label{deltaPhi1}
\end{equation}
and the Bianchi equations $B_1$ and $B_2$:
\begin{equation}
\label{DPsi1}
D\, \Psi_{{1}} =4\,\rho\,\Psi_{{1}} ,
\end{equation}
\begin{equation}
\label{deltaPsi1}
\delta \, \Psi_{{1}} =2 \left( 2\, \bar{\alpha} +3 \,\beta\right) \Psi_{{1}} -3\,\sigma\,\Psi_{{2}}-2\,\sigma\,\Phi_1 \,\bar{\Phi_1} .
\end{equation}
\par
Next, we calculate $[\delta , D ] \, \Psi_1$, which by (\ref{DPsi1}), (\ref{deltaPsi1}) and the commutator relation (\ref{com14}) leads to the expression :
\begin{eqnarray}
\left( \delta \,\rho-\bar{\alpha}\,\rho-3 \sigma\,\alpha -\beta\,\rho-\bar{\beta }\, \sigma -\frac{3}{2} {\Psi_{{1}}} \right) \Psi_{{1}}+\sigma\,\bar{\delta}\,\Psi_{{1}} +2 \sigma\,\rho\,\,\Phi_1\,\bar{\Phi_1} =0.
\label{temp1}
\end{eqnarray}
Applying the $D$-operator to this equation gives :
\begin{eqnarray}
\nonumber
 & \sigma\,{D} \,\bar{\delta}\, \Psi_{{1}} + \Psi_1\,{D} \,\delta \,\rho +2\, \rho\,\sigma\,\bar{\delta}\,\Psi_{{1}} +4\,\rho\,\Psi_1\,\delta\,\rho+ \left( 2\, {\sigma}^{2}+14\,{\rho}^{2} \right)\sigma\, \Phi_1\,\bar{\Phi_1} \\
&-\left( 6\left( \bar{\alpha } +\beta \right) {\rho}^{2} + \left( 22\sigma\,\alpha +8\bar{\beta }\, \sigma +13{\Psi_{{1}}}\right) \rho +\sigma\,\bar{\Psi_{{1}}}  +2\left( \bar{\alpha}+2\beta \right){\sigma}^{2} \right) \Psi_{{1}} \nonumber\\
 &= 0
\label{temp2}
\end{eqnarray}
The two expressions (\ref{temp1}) and (\ref{temp2}) enable us to simplify $[\bar{\delta}, D](\rho\,\Psi_1)$ :
\begin{eqnarray}
\nonumber
\left( 5\left(\alpha+ \bar{\beta} \right) {\rho}^{2} - \left( \left( \bar{\alpha} +\beta \right)\sigma +5 \bar{\delta} \, \rho +2\,\delta \,\sigma -\bar{\Psi}_{{1}}\right)\rho  \right)\Psi_1 \\
+\frac{5}{2}\,{\frac {{\Psi_{{1}}}^{2}{\rho}^{2}}{\sigma}}+3\,{\sigma}^{2}\Psi_{{2}}\rho=0 .
\label{tt}
\end{eqnarray}
Eliminating $\bar{\delta}\,\rho$ by $NP_{11}$ allows us to solve for $\delta \, \sigma $ :
\begin{equation}
\label{deltasigma}
\delta\, \sigma  = \frac {5}{14}\,\frac {\rho\,\Psi_{{1}}}{\sigma}+\frac{3}{7}\,{\frac {{\sigma}^{2}\Psi_{{2}}}{\Psi_{{1}}}}+2\,\sigma\,\bar{\alpha}-\frac{6}{7} \,\beta\,\sigma+\frac{6}{7}\,\bar{\Psi}_{{1}}.
\end{equation}
Notice that $\Psi_1$ has to be non-zero, but as we are assuming nonvanishing shear, this condition is fulfilled automatically : if $\Psi_1 = 0$, then $\Phi_1 =0$ (by (\ref{temp1})) and hence also $\Phi_2 = 0$ by (\ref{deltaPhi1}). 

\par
If we substitute (\ref{deltasigma}) in (\ref{tt}) we can solve for $\delta\,\rho $,
and by substituting the result in (\ref{temp1}) we get an equation for $\bar{\delta} \, \Psi_1 $ :
\begin{equation*}
\label{deltarho}
\delta \, \rho = \frac{3}{7} \,{\frac {{\sigma}^{2} \bar{\Psi}_{{2}}}{\bar{\Psi}_{{1}}}}+{\frac {5}{14}}\,{\frac {\bar{ \Psi}_{{1}} \rho}{\sigma}}-\sigma\,\alpha + \frac{1}{7}\,\bar{\beta} \sigma-\frac{1}{7}\,\Psi_{{1}}+\rho\,\beta+\rho\,\bar{\alpha},
\end{equation*}
\begin{equation*}
 \label{deltabarPsi1}
 \bar{\delta} \, \Psi_{1} = \left( {\frac {23}{14}}\,{\frac {{\Psi_{{1}}}}{\sigma}} -\frac{3}{7} \,{
\frac {\sigma\,\bar{\Psi}_{{2}} }{\bar{ \Psi}_{{1}} }}+4\,\alpha+\frac{6}{7}\,\bar{ \beta} -{\frac {5}{14}}\,{\frac {\rho\,\bar{ \Psi}_{{1}}}{{\sigma}^{2}}} \right) \Psi_{{1}}-2\,\Phi_1\,\rho\,\bar{\Phi_1} .
\end{equation*}
The next step in our calculation is to look at $B_3$, which by now can be solved for $D \, \Psi_2$ :
\begin{equation*}
\label{DPsi2}
{D} \, \Psi_{{2}} =-{\frac {5}{14}}{\frac {\Psi_{{1}}\bar{\Psi}_{{1}} \rho}{{\sigma}^{2}}}-\frac{3}{7}\,{\frac {\sigma\,\Psi_{{1}}\bar{\Psi}_{{2}}}{\bar{\Psi}_{{1}}}}+2\alpha\Psi_{{1}}+\frac{6}{7}\Psi_{{1}}\bar{\beta} +{\frac {23}{14}}{\frac {{\Psi_{{1}}}^{2}}{\sigma}}+3\rho\Psi_{{2}}.
\end{equation*}
Substituting all the above in $[\delta , D] \, \rho $, we find an expression for $\beta$ :
\begin{equation*}
\beta = \frac{1}{8} \, \frac{4\, \Psi_2 \sigma^3 + \Psi_1 \sigma \bar{\Psi_1} + \Psi_1^2 \rho}{\Psi_1 \sigma^2}.
\end{equation*}
This almost finishes our proof. The next steps are to look at $[\delta , D] \, \Phi_1$ and Maxwell's equation (\ref{max2}) which can be solved for $\bar{\delta} \, \Phi_1$ and $D \,\Phi_2$ after which $[\bar{\delta}, D] \,\Phi_1$ yields an expression for $\Phi_2$ :
\begin{eqnarray*}
\bar\delta \, \Phi_1  =-\frac{1}{4} \,{\frac {\Phi_1 \,\bar{\Psi}_{{1}} \rho}{{\sigma}^{2}}}+2\,\Phi_1\,\alpha+\frac{5}{4}\,{\frac {\Phi_1 \,\Psi_{{1}}}{\sigma}}, \\
{D} \, \Phi_2 =\rho\,\Phi_2-\frac{1}{4}\,{\frac {\Phi_1\,\bar{ \Psi}_{{1}} \rho}{{\sigma}^{2}}}+2\,\Phi_1\,\alpha+\frac{5}{4}\,{\frac {\Phi_1\,\Psi_{{1}}}{\sigma}}, \\
\Phi_2=\frac{1}{2}\,{\frac {\Phi_1\, \left( -{\Psi_{{1}}}^{2}\rho+2\,\Psi_{{2}}{\sigma}^{3} \right) }{{\sigma}^{3}\Psi_{{1}}}}.
\end{eqnarray*}
We then evaluate $[\bar{\delta}, \delta]\, \Phi_1$, from which we eliminate $\bar{\delta}\,\bar{\alpha}$ by $[\bar{\delta}, \delta] \, \Psi_1$. This results in the expression (\ref{vgl1}) below. We also write down the equation for $[\bar{\delta}, \delta] \, \Psi_1 $, after eliminating  from it $\delta\,\alpha$ by $[\bar{\delta}, \delta] \, \sigma $ (namely (\ref{vgl2})).
\begin{eqnarray}
\nonumber
3\,\rho \,\sigma{\Psi_{{1}}}^{4}\bar{\Psi_1} -4\left(\,{\sigma}^{2} +2\,\rho^2\right) {\Psi_{{1}}}^{3}\bar{\Psi_1} ^{2} +\rho\, \sigma {\Psi_{{1}}}^{2}\bar{\Psi_1}^{3} -24\,\bar{\alpha} \, {\sigma}^{3}{\Psi_{{1}}}^{3}\bar{\Psi_1} \\
\nonumber
-8\,\alpha{\sigma}^{3} {\Psi_{{1}}}^{2}\bar{\Psi_1}^{2} +12\,\rho\,{\sigma}^{3}\bar{\Psi_2}{\Psi_{{1}}}^{3} +8 \left( 7\,{\sigma}^{2} -4\,{\rho}^{2}\right){\sigma}^{2}\,\Phi_1\,\bar{\Phi_1}  {\Psi_{{1}}}^{2} \bar{\Psi_1}\\
+4 \left( \Psi_{{2}}-2\Phi_1 \right) \rho {\sigma}^{3}\Psi_{{1}}\bar{\Psi_1}^{2} +32\left( 2 \alpha \Psi_{{1}} -  \rho \Psi_{{2}}\right) {\sigma}^{5} \Phi_1 \bar{\Phi_1}\bar{\Psi_1} =0, \label{vgl1}
\end{eqnarray}
\begin{eqnarray}
\nonumber
-\rho\, \sigma\,{\Psi_{{1}}}^{4}\bar{\Psi_1} + 4\left( {\sigma}^{2} +2\,{\rho}^{2} \right) {\Psi_{{1}}}^{3}\bar{\Psi_1}^{2}-3\,\rho\,\sigma \,{\Psi_{{1}}}^{2} \bar{\Psi_1}^{3}
+8\,\bar{\alpha}\,{\sigma}^{3}{\Psi_{{1}}}^{3}\bar{\Psi_1} \\
\nonumber
+ 24\,\alpha \,{\sigma}^{3}{\Psi_{{1}}}^{2}\bar{\Psi_1} ^{2} -4\,\rho\,{\sigma}^{3}\bar{\Psi_2} {\Psi_{{1}}}^{3}+8\left( 4\,{\rho}^{2} -{\sigma}^{2} \right){\sigma}^{2}\Phi_1\,\bar{\Phi_1}{\Psi_{{1}}}^{2} \bar{\Psi_1} \\
-4\left( 3\,\Psi_{{2}}+2\,\Phi_1\,\bar{\Phi_1}\right)\rho\,{\sigma}^{3}\Psi_{{1}}\bar{\Psi_1} ^{2}
+32\left( 2\,\alpha \, \Psi_{{1}}-\,\rho\,\Psi_{{2}}\right){\sigma}^{5}\Phi_1\,\bar{\Phi_1} \bar{\Psi_1} =0. \label{vgl2}
\end{eqnarray}
Notice that we can solve these equations for $\alpha$ and $\bar{\alpha}$,
\begin{eqnarray}
\label{alpha}
\alpha=\frac{1}{8}\,{\frac {-2\,{\Psi_{{1}}}^{2}{\rho}^{2}+4\,\Psi_{{2}}\rho\,{\sigma}^{3}-{\Psi_{{1}}}^{2}{\sigma}^{2}+\rho\,\Psi_{{1}}\sigma\,\bar{\Psi}_{{1}} }{{\sigma}^{3}\Psi_{{1}}}}, \\
\label{alphabar}
\bar{\alpha } =\frac{1}{8}{\frac {-2\,{\bar{\Psi_{{1}}}}^{2}{\rho}^{2}+4\,\bar{\Psi_{{2}}}\rho\,{\sigma}^{3}-{\bar{\Psi_{{1}}}}^{2}{\sigma}^{2}+\rho\,\Psi_{{1}}\sigma\,\bar{\Psi_{{1}}}}{{\sigma}^{3}\bar{\Psi_{{1}}}}}+2{\frac {\Phi_1\,\bar{\Phi_1}  \left( {\sigma}^{2}-{\rho}^{2} \right) }{\Psi_{{1}}\sigma}},
\end{eqnarray}
but, as can easily be seen from the above results,
this implies
\begin{equation*}
2\,{\frac {\Phi_1\,\bar{\Phi_1}  \left( {\sigma}^{2}-{\rho}^{2} \right) }{\Psi_{{1}}\sigma}}=0,
\end{equation*}
which shows that we are in the cylindrical class, and thus proves our statement.

\section{Conclusion}
We have shown that, in the presence of an aligned Maxwell field, no generalization of the `spherical' Newman Tamburino solutions exists. As the final equation shows, applying such a Maxwell field forces us into the cylindrical class. The next step in our investigation will be to try to integrate the latter. Although the calculations are rather more lengthy and complicated than in vacuum, there is good hope that this class will be fully integrable.

\section{Appendix}
\subsection*{The commutators:}
\begin{equation} \begin{array}{lll}
\label{com34}
\left( \Delta D - D \Delta \right) &=& \left( \gamma + \bar{\gamma} \right) D + \left( \epsilon + \bar{\epsilon} \right) \Delta - \left( \tau + \bar{\pi} \right) \bar{\delta} - \left( \bar{\tau} + \pi \right) \delta, \\
\label{com14}
\left( \delta D - D \delta \right) &=& \left( \bar{\alpha} + \beta - \bar{\pi} \right) D + \kappa \Delta - \sigma \bar{\delta} - \left( \bar{\rho} + \epsilon - \bar{\epsilon} \right) \delta, \\
\label{com13}
\left( \delta \Delta - \Delta \delta \right) &=& -\bar{\nu} D + \left( \tau - \bar{\alpha} - \beta \right) \Delta + \bar{\lambda} \bar{\delta} + \left( \mu - \gamma + \bar{\gamma} \right) \delta, \\
\label{com21}
\left( \bar{\delta} \delta - \delta \bar{\delta} \right) &=& \left( \bar{\mu} - \mu \right) D + \left( \bar{\rho} - \rho \right) \Delta - \left( \bar{\alpha} - \beta \right) \bar{\delta} - \left( \bar{\beta} - \alpha \right) \delta.
\end{array}
\end{equation}

\section*{Acknowledgment}
 NVdB expresses his thanks to Robert Debever($\dagger$) 
 and Jules Leroy (Universit\'e Libre de Bruxelles), whose questions about the Newman Tamburino metrics have led to the present investigation.

\section*{References}


\begin{thebibliography}{1}
\expandafter\ifx\csname url\endcsname\relax
  \def\url#1{{\tt #1}}\fi
\expandafter\ifx\csname urlprefix\endcsname\relax\def\urlprefix{URL }\fi
\providecommand{\eprint}[2][]{\url{#2}}

\bibitem{Newman1}
Newman E and Tamburino L 1962 {\em J. Math. Phys.\/} {\bf 3} 902--907

\bibitem{Kramer}
Kramer D, Stephani H, MacCallum M~A~H, Hoenselaers C and Herlt E 2003 {\em
  Exact solutions of Einstein's field equations\/} (Cambridge University Press)

\end{thebibliography}
\providecommand{\newblock}{}

\end{document}